# Analyzing animal movement using deep learning

Thibault Fronville[1,2,*] & Maximilian Pichler[3,*], Johannes Signer[4], Marius Grabow[1], Stephanie Kramer-Schadt[1,2], Viktoriia Radchuk[1,†] & Florian Hartig[3,†]

[1] Leibniz Institute for Zoo and Wildlife Research (IZW), Department of Ecological Dynamics, Alfred-Kowalke-Straße 17, 10315 Berlin, Germany

[2] Technische Universität Berlin, Institute of Ecology, Rothenburgstr. 12, 12165 Berlin, Germany

[3] University of Regensburg, Theoretical Ecology, Universitätsstraße 31, 93053 Regensburg, Germany

[4] University of Goettingen, Wildlife Sciences - Faculty of Forest Sciences and Forest Ecology, Büsgenweg 3, 37077 Göttingen, Germany

Corresponding Authors: radchuk@izw-berlin.de, florian.hartig@biologie.uni-regensburg.de

## Abstract

1. Understanding how animals move through heterogeneous landscapes is central to ecology and conservation. In this context, step selection functions (SSFs) have emerged as the main statistical framework to analyze how biotic and abiotic predictors influence movement paths observed by radio tracking, GPS tags, or similar sensors.

2. A traditional SSF consists of a generalized linear model (GLM) that infers the animal's habitat preferences (selection coefficients) by comparing each observed movement step to a number of random steps. Such GLM-SSFs, however, cannot flexibly consider non-linear or interacting effects of predictors on the movement, unless those have been specified *a priori*. To address this problem, generalized additive models (GAMs) have recently been integrated in the SSF framework, but those GAM-SSFs are still limited in their ability to represent complex habitat preferences and inter-individual variability.

3. Here we explore the utility of deep neural networks (DNNs) to overcome these limitations. We find that DNN-SSFs, coupled with explainable AI (xAI) to extract selection coefficients, offer many advantages for analyzing movement data. In the case of linear effects, they effectively retrieve the same effect sizes and p-values as conventional GLMs. At the same time, however, they can automatically detect complex interaction effects, nonlinear responses, and inter-individual variability if those are present in the data.

4. We conclude that DNN-SSFs are a promising extension of traditional step selection models. Our analysis extends previous research on DNN-SSF by exploring differences and similarities of GLM, GAM and DNN-based SSF models in more depth, in particular regarding the validity of statistical indicators such as p-values and confidence intervals that are derived from the DNN. We also propose new DNN structures to capture inter-individual effects that can be viewed as a nonlinear random effect. All methods used in this paper are available via the 'citoMove' R package.

* Shared first authorship
† Shared last authorship

# Introduction

Animal movement is a fundamental ecological process that is the basis for a wide range of ecological and evolutionary phenomena. Understanding how, when, and why animals move is pivotal for predicting individual survival and reproductive success, and for explaining their spatial distribution and thus population and ecosystem dynamics (Bowler and Benton 2005; Nathan et al. 2008; Jeltsch et al. 2013; Bauer and Hoye 2014; Schlägel et al. 2019). Developing analytical methods to explore how animal movement emerges in interaction with the environment is therefore crucial for ecological research and conservation efforts (Nathan et al. 2008).

The most common framework to analyze animal movement is the step selection function (SSF) approach (Avgar et al. 2016; Fieberg et al. 2021). SSF models analyze how an animal selects its next movement step by comparing the observed trajectory to all other possible movement options, or to a limited set of predefined alternatives (see integrated SSF, Avgar et al. 2016; Fieberg et al. 2021). Mathematically, this corresponds to evaluating the relative likelihood of observed movement steps by integrating over all possible movement steps. This use-availability design at the scale of each movement step makes it possible to quantify interactions of the animal with its biotic and abiotic environments.

The traditional approach to fit an SSF model is to approximate the theoretically continuous kernel of all possible movement steps by a finite number of randomly sampled potential steps. The contrast between these random steps and the observed steps is then fit using a conditional logistic regression, a generalized linear model (GLM) with a logit link (Therneau 2024; Signer et al. 2019; but other approaches exist, see Michelot et al. 2024; Muff et al. 2020). The creation of random steps as a contrast is similar to pseudo-absences in species distribution models, except that in a SSF model, a separate set of random steps is created exclusively for each observed step. These random steps are usually based on observed turning angles and step lengths to separate habitat selection and movement processes (Fieberg et al. 2021). This approach, which we refer to as a GLM-SSF, has been extensively applied in the analysis of empirical movement data, for example, to study animals' selective preference towards landscape features (Thurfjell et al. 2014), anti-predator behaviors (Latombe et al. 2014), human-wildlife interactions (Bjørneraas et al. 2011) or inter-individual interactions (Schlägel et al. 2019; Dickie et al. 2020).

A limitation of a GLM-SSF, as of any other parametric statistical model, is that the functional relationship describing the influence of biotic and abiotic conditions on the movement must be specified *a priori*. In practice, GLM-SSFs typically rely on linear and quadratic effects for all predictors. The real response of animals, however, is often more complex and likely to display nonlinear relationships or interactions between predictor variables (Mysterud and Ims 1998;

Avgar et al. 2016). For instance, an animal might prefer moderate vegetation cover for foraging but avoid both open areas (due to predation risk) and dense vegetation (which impedes movement), resulting in a nonlinear, bell-shaped selection pattern. Also, habitat preferences are not always constant over time (Richter et al. 2020; Forrest et al. 2025a), and might shift, for example, between the seasons (Boyers et al. 2019). Moreover, even when the random steps are sampled from the observed movement distribution, it may be necessary to still include movement variables in the model, because habitat structure can modify the movement decisions (iSSF, see Avgar et al. 2016; Fieberg et al. 2021). While it is in principle possible to represent such complex relationships in a GLM-SSF, it is difficult to anticipate the functional form of these patterns *a priori*, motivating the search for modelling approaches that more flexibly adjust to the observed data.

Movement ecology  research in recent years has therefore focused on finding more flexible modelling approaches, for example by including time-varying selection patterns (Richter et al. 2020), incorporating multiple movement types (e.g. foraging and migration) to reflect behavioral variation (Klappstein et al. 2024) or accounting for inter-individual heterogeneity in habitat selection (Chatterjee et al. 2024; Klappstein et al. 2024; Muff et al. 2020). These extensions have been implemented using a range of statistical frameworks, including generalized additive models (GAMs), mixed-effects models, and state-dependent (e.g. hidden Markov) models. Accounting for inter-individual variability in the SSF is a particularly interesting problem because animals' habitat selection often varies strongly through individual differences or personality (e.g., shy vs. bold behavior; Sloan Wilson et al. 1994). Ignoring such inter-individual variability can result in biased selection estimates (Lesmerises and St-Laurent 2017). To capture inter-individual variability and more complex habitat selection patterns, Klappstein et al. (2024) integrated random slopes and hierarchical smoothing terms (GAMs) into SSFs (GAM-SSFs). While these models offer more flexibility than GLM-SSFs, they are still limited in some dimensions, particularly when including a large number of nonlinear interactions (see, e.g., Pedersen et al. 2019) or when considering interactions between habitat selection and movement preferences.

To allow for a more flexible approach to modelling animal movement, a logical next step is to follow the general trend in ecology towards flexible machine learning (ML) and deep learning (DL) algorithms as an alternative to the more rigid classical models from parametric statistics (Christin et al. 2019; Borowiec et al. 2022; Pichler and Hartig 2023b). ML is already being used in movement ecology to classify behavioral states from their movement (Christin et al. 2019; Wang 2019) or species from images (Tabak et al. 2019); predict movement trajectories (Torney et al. 2021); or analyze habitat use (Cífka et al. 2023; Forrest et al. 2025b). For example, ML has offered the possibility to classify animals' activities such as feeding and resting from analyzing camera traps images (Norouzzadeh et al. 2018). Another example is Browning et al. (2018), who predicted the diving behaviors of seabirds from GPS data with deep neural networks (DNN).

Consistent with this development, some recent studies successfully integrated ML and DL approaches within the SSF framework (Wijeyakulasuriya et al. 2020; Cífka et al. 2023; Forrest et al. 2025b). We consider DNNs to be a particularly promising approach for this purpose, as they naturally account for nonlinear, high-dimensional (higher order interactions between variables) relationships while being able to retrieve near-unbiased parameter estimates (Pichler and Hartig 2023a). Furthermore, they are highly extensible and can be adapted to more advanced DL architectures. For example, Cífka et al. (2023) employed a Transformer-based movement model (MoveFormer) that considers past movement steps in the next movement decision, and Forrest et al. (2025b) used a convolutional neural network (CNN) architecture (deepSSF) that is capable of processing complex inputs such as images/sounds (Forrest et al. 2025b).

These first results highlight the promise of DNN for SSF models, but important questions remain that hinder their wider application for analyzing animal movement. A first fundamental question is whether DNNs, which are clearly more flexible, but potentially also data-hungry, work well with comparatively small datasets that are typical for wildlife ecology. Second, neural networks do not natively provide effect sizes and p-values, which is crucial for ecologists who want to interpret the fitted models to understand the mechanisms behind animal movement. Explainable AI (xAI) techniques can make these "black-box" models more interpretable (Ryo et al. 2021) and even provide p-values and confidence intervals. However, since ML models often trade bias against variance in parameter optimization (Shmueli 2010), it is somewhat unclear whether these extracted effects and their statistical validity are reliable in situations that are typical for wildlife ecology. And thirdly, DNNs have a reputation for being difficult to use and less well supported in the R environment than comparable methods such as GLMs or GAMs, which arguably prevents uptake of these new methods in the movement ecology community.

Here, we address all three challenges by comparing the performance of DNNs with GLM and GAM approaches for understanding the drivers of animal movement. Unlike other previous studies that introduced neural networks in the SSF framework, we concentrate on fully connected DNN (also known as multi-layer perceptrons or MLPs). While these are considered the simplest network architectures, they are also closest in structure and inputs to the more established algorithms such as GLMs or GAMs. Moreover, for MLPs, it is relatively straightforward to extract effect sizes and p-values, which allows us to compare their inferential performance with the established models in a typical ecological analysis, rather than comparing only their predictive performance.

Using simulated data, our analysis focuses on the key challenges commonly encountered in real-world movement data: (i) inferring correct p-values and uncertainties for linear effects, (ii) capturing nonlinear patterns of habitat selection, (iii) revealing complex interactions between environmental predictors, and (iv) accounting for inter-individual variability in selection behavior. We finally applied DNN-SSFs to a case study (based on Stillfried et al. 2017) with tracking data from wild boar inhabiting urban and rural areas. Here, our DNN-SSFs were able to infer biologically plausible differences in movement behavior that correlated with individuals' home ranges.

## Methods

## The step selection function (SSF) framework

The idea of the step selection function (SSF; Fortin et al. 2005; Rhodes et al. 2005) is to build a statistical model that compares the decision of the animal taken at each movement step to its (infinite) possible alternative movement steps.

In the traditional SSF model, these alternatives were defined based on general characteristics of the movement process, such as average step length and turning angle. In practice, this was done by sampling, for each observed step, several random steps from a movement kernel fitted to the observed movement steps. The sets of observed/available steps form groups that are called strata, with each stratum representing a single decision point in the animal's movement. This approach effectively factors out the general movement characteristics of the animal through the random sample, resulting in a model with habitat selection effects only. The downside of this approach, however, is that the observed movement characteristics (e.g. step length) may be affected by the structure of the habitat. Thus, the approach can create biased estimates if the habitat structure substantially influences the general movement characteristics.

This limitation was later addressed by Avgar et al. (2016), who proposed an extension of the SSF approach that integrates movement behavior and habitat selection into a unified framework, now commonly referred to as integrated step selection functions (iSSFs). A common way to implement this model is via a conditional logistic regression with a compound likelihood:

$$p(s_{t+1}|s_t) = \frac{w(s_{t+1})\phi(s_{t+1},s_t)}{\int_\Omega w(z)\phi(z|s_t)dz} \quad \text{equation (1)}$$

$$w(s_{t+1}) = \exp(\beta_1 x_1(s_{t+1}) + \beta_2 x_2(s_{t+1}) + \cdots + \beta_n x_n(s_{t+1})) \quad \text{equation (2)}$$

where $p$ is the probability of moving to a location $s_{t+1}$ given the previous location $s_t$; $w$ is the exponential (log-linear) selection function that describes the effects $\beta_n$ of the environmental covariates $x_n$ at location $s_t$; and $\phi$ is a selection-independent movement kernel that reflects the movement patterns of the animal in a homogenous landscape. In practice, $\phi$ is often expressed as a function of the step length and turning angle. The iSSF model thus understands movement as a compound process where the animal has a certain general movement behavior (described by the kernel $\phi$) which is modified by habitat preferences (described by the selection effects w).

In practice, evaluating the likelihood in eq.1 requires integrating over all possible steps the animal could have taken ($\Omega$). As this integral typically has no closed-form solution, it is typically approximated using Monte-Carlo integration, i.e. by randomly sampling a finite set of possible steps for each observed step. For numeric efficiency, these random steps are usually sampled non-uniform according to the empirically observed distributions of step lengths (e.g., Gamma distribution) and turning angles (e.g., von Mises distribution), as in the normal SSF. In this case, however, the parameters of the fitted movement kernel $\phi$ have to be corrected later in the likelihood or in the parameters to account for the non-uniform sampling (Michelot et al. 2024). A certain downside of this approach is that it makes modelling statistical interactions between the movement kernel $\phi$ and the selection function $w$ technically challenging, a problem which will be relaxed by our DNN-SSF later.

This traditional parametric conditional logistic SSF model is implemented in several statistical tools, among them the "amt" (Signer et al. 2019) or "survival" (Therneau and Grambsch 2000) packages in R. As discussed in the introduction, a crucial limitation of these parametric models is that they require the user to specify the functional relationships that are assumed *a priori*. This requirement is somewhat relaxed by the possibility to implement SSFs with GAMs through the R package "mgcv" (Wood 2011; for more details see Klappstein et al. 2024), which can flexibly fit functional relationships, but still require explicit specification of other structures, in particular interactions. The goal of finding a more flexible model motivates the use of DNNs in the SSF framework.

## Introducing DNNs in the step selection framework

As discussed in the introduction, in theory, it is straightforward to replace the parametric movement and selection functions by a neural network. In practice, however, few studies have tested this approach so far. A first problem for assessing the performance of a DNN-SSF is that, unlike for GLM-SSF and GAM-SSF, there is so far no R package available for fitting these models "off-the-shelf". It is of course possible to create custom implementations of DNN-SSF in the existing DL frameworks in R or python, as done in previous studies (e.g. Cífka et al. 2023; Forrest et al. 2025b), but this approach makes re-use tedious and thus limits the utility of the methodological insights generated for ecological practitioners.

For general neural networks, the 'cito' R package (Amesoeder et al. 2024a) aims at solving the problem of making these models available to a wide range of users. 'cito' provides an R interface to 'torch', a state-of-the-art deep learning framework (Falbel and Luraschi 2025). The 'cito' model specification philosophy follows the structure of many popular R regression packages. Specifically, it uses the R formula syntax to specify the predictors and offers the ability to switch between different families and link functions. Moreover, cito provides many downstream functions, in particular explainable AI (xAI) methods to extract or plot the effect of predictors on model predictions. However, cito does not provide support for the conditional logit link, which would allow using it for the analysis of animal movement.

To solve this problem, we developed a wrapper that extends 'cito' to DNN-SSFs. To explain the functioning of this new 'citoMove' package, remember that, in a traditional SSF, the probability of selecting a step to location $s_{t+1}$, given the environmental covariates $x_i$ at that location (see eq. 1), depends on

$$p(s_{t+1}|s_t) \sim w\big(x_{s_{t+1}}\big)\phi(s_{t+1}, s_t) \quad \text{equation (3)}$$

To enable more flexible modeling, we replace the right side of the equation by a neural network parameterized by weights $\theta$:

$$p(s_{t+1}|s_t) \sim \exp(f_\theta(\mathrm{x}_{t+1}, \mathrm{s}_{t+1}, s_t)) \quad \text{equation (4)}$$

The training of these weights is then delegated to the 'cito' R package. Thus, 'citoMove' simply replaces the parametric specifications of both the selection function and the movement kernel in a conventional GLM-SSF with a flexible neural network, while everything else (in particular the link function and the likelihood) stays identical. Interestingly, this approach that integrates the movement kernel $\phi$ and the selection function $w$ into a single function naturally also considers interactions between movement and habitat selection.

A disadvantage of using a DNN to model the animals' response to the predictors is that a DNN does not directly provide interpretable effect estimates and associated uncertainties such as p-values and confidence intervals. However, effect estimates and variable importance can be retrieved post-hoc, for example by exploring the (average) response of the fitted model to changes in the inputs. There are a large number of methods and metrics available for this purpose, collectively known as explainable AI (xAI, e.g. Ryo et al. 2021; Holzinger et al. 2022; Dwivedi et al. 2023). For 'citoMove', we make use of the fact that cito already calculates several xAI metrics, most importantly permutation importance, average conditional effects (Molnar et al. 2020; Pichler and Hartig 2023a), and accumulated local effect plots (Apley and Zhu 2020).

A certain complication of DL models, which may also be seen as an advantage, is that they have many hyperparameters that control the complexity of the learned input-output relationship. These hyperparameters allow adjusting the complexity of the fitted relationships to the specific application; however, it also means that these hyperparameters should be optimized, typically by varying them and observing model performance on an independent hold-out dataset. Although we did not use these features in our current analysis, 'cito' aids the modeler in this task through several built-in functions for hyperparameter tuning.

## Comparison of DNN-SSF to GLM-SSF and GAM-SSF

Having developed a DNN-SSF with essentially the same usability and functions as the more traditional SSFs based on GLMs or GAMs, we proceeded with five experiments to compare the performance of the three modeling approaches in simulated scenarios where the true underlying movement process is perfectly known (Figure 1). The first three experiments compare performance of the three models in a) estimating linear effects (including p-values) of a single linear predictor (Figure 1, scenario 1), b) estimating nonlinear effects of single predictors (Figure 1, scenario 2), and c) estimating statistical interactions of multiple predictors (Figure 1, scenario 3). The last two experiments do not have a direct counterpart in GLMs or GAMs, but highlight unique advantages of DNN-SSFs, namely d) the assessment of inter-individual variability in selection (similar to a random intercept effect in GLMMs or hierarchical GAMs; Figure 1, scenario 4), and e) the assessment of dynamically changing predictors, like distances between moving individuals (Figure 1, scenario 5).

| Scenario Description | Visualization | Purpose |
|---|---|---|
| **1. Linear effect**<br>Linear selection for one predictor x | 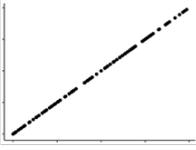 | Compare DNNs to GLM-SSFs in inferring p-values and uncertainties |
| **2. Nonlinear effect**<br>Nonlinear selection for one predictor x | 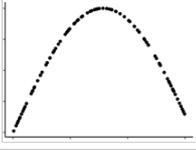 | Compare DNNs to GAM-SSFs in inferring nonlinear effects |
| **3. Statistical interactions**<br>Linear selection for multiple predictors x(n) that are also in pairwise interactions<br>Predictors : 9<br>Pairwise interactions : 3 | 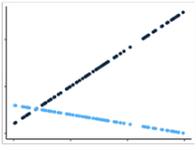 | Compare DNNs to GLM-SSFs in inferring multiple pairwise interaction effects |
| **4. Embeddings -**<br>**Variability in focal individual**<br>Linear selection for multiple predictors xn that varies by group (e.g. species, ID, sexes)<br>Predictors : 20<br>Individuals : 20 ; Groups : 4 | 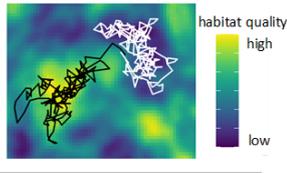 | Check whether DNNs are able to infer inter-individual variability in the response of the focal individual |
| **5. Embeddings -**<br>**Variability in opponent individual**<br>Linear selection for distance to opponent that varies by group of opponent (e.g. species, ID, sexes)<br>Predictors : 1<br>Individuals : 15; Opponent Groups : 3 | 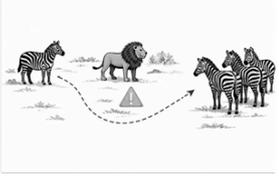 | Check whether DNNs are able to infer inter-individual variability in the effect of the opponent individual |

**Figure 1: Summary of the five scenarios built to test the inferential performance of DNN-SSFs.** The first three scenarios were built to compare the inferential performance of DNN-SSFs to existing methods (GLM-SSF & GAM-SSF). The 4th and 5th scenarios were built to explore the performance of DNN-SSFs in inferring inter-individual variability in habitat selection and dynamically changing predictors.

Our first scenario was built to test the inferential performance of DNNs in recovering a linear effect of a single environmental predictor with correct statistical properties (5% type I error rate of the p-values, nominal coverage of the confidence intervals, unbiased parameter estimates). We calculated p-values and confidence intervals for the DNN-estimated effects based on 20 bootstrap replicates. We compared the results to a GLM-SSF only, as this model is theoretically the best choice if one is certain that effects are linear. Due to its higher flexibility, we expected that the DNN-SSF would perform slightly worse than the GLM-SSF (as predicted by the bias/variance trade-off). The goal of the experiment was, however, to show these performance losses are moderate.

For the second scenario, we tested the DNNs' ability to infer a nonlinear effect of one predictor. We compared the inference of DNN-SSFs to the inference obtained with GAM-SSFs using the 'mgcv' package (Wood 2025) with default spline penalties, which we see as the state-of-the-art choice for recovering univariate nonlinear effects. We expect that DNN-SSFs can infer nonlinear responses similarly to GAM-SSFs.

For our third scenario, we tested the performance of DNNs in inferring statistical interactions among multiple predictors. For each effect, we ran 100 repetitions. From the estimated slopes, we then calculated the p-value, mean squared error (MSE), and variance. This scenario serves to demonstrate advantages of DNNs as flexible models in higher-dimensional settings, by adding multiple predictors that are also in pairwise interactions and assess whether DNNs can reveal non-additive relationships that GLM-SSFs might miss.

In our fourth (d) and fifth (e) scenario, our goal was not the comparison to established models but investigating a feature of DNNs that does not have a direct counterpart in GLMs or GAMs. This feature arises in the context of categorical grouping variables such as individuals or years. In a DNN, these can enter the model directly as any other variable and thus moderate the effect of environmental predictors corresponding to normal interactions in a GLM, but it is also possible to implement a model architecture where categorical predictors are first embedded into a lower-dimensional numeric space before they are used in the larger network (Guo and Berkhahn 2016). The optimal number of embedding dimensions depends on the complexity of the underlying hierarchical structure of the predictor and can be tuned. Similar to an ordination, the position in the embedding space corresponds to how similar individuals are in their environmental responses (e.g., habitat selection). To interpret the embedding space, we can back-project how environmental effects are dependent on the embedding positions, akin to a bi-plot in an ordination. The embedding structure may be used to capture inter-individual variability, similar to a random intercept and random slope structure in mixed-effects models (GLMMs or GAMMs) or in hierarchical GAMs (Pedersen et al. 2019), but with the advantage that this extends to all kind of similarities (also of individual interactions) in a nonlinear way while still having a relatively simple visual interpretation.

In our fourth scenario (Figure 1), we tested the ability of this embedding architecture to represent and infer inter-individual variability in habitat selection. To do that, we simulated 20 individuals divided into four groups, which can be thought of as four personality types (five individuals per group) that react to an environment with 20 predictors. Individuals of the same group respond similarly to the environmental predictors. We designed the simulation such that five of the 20 environmental predictors did not affect the step selection of any individual at all ($\beta = 0$). Among the 15 predictors with effect, coefficients for 10 environmental predictors were randomly generated ($\beta > -3$ & $\beta < 3$) and the other five environmental predictors were correlated ($\beta > -3$ & $\beta < 3$). The reason for the setting was that we expected correlated and uncorrelated predictors to manifest differently in the embedding space which can be seen in our results. For this scenario, one hierarchical level (with two embedding dimensions) is present (grouped individuals).

In the fifth scenario (Figure 1), we consider the problem that individuals may not only differ in their response to the environment, but also in their effect on other individuals. Specifically, we consider three groups of five individuals each that move together in a homogenous environment. At each time step, the nearest other individual would act as an 'opponent' to influence the movement of a focal individual ('social' predictor). Depending on the type of the opponent, focal individuals are either repelled, attracted or respond neutrally to the opponent (these responses are referred to as "individual interactions"). This example is designed to simulate a social structure where individuals' movement preferences vary depending on the specific individuals in their surroundings. For this scenario, one hierarchical level (with two embedding dimensions) is present (grouped opponents).

For all scenarios, simulation settings were chosen to reflect sample sizes typical for empirical movement data. The DNN models were trained for 100 to 150 epochs (training times). We confirmed that this was enough to reach stable convergence in all scenarios. As the DNNs were relatively simple, training the weights took only seconds to minutes per network, depending on dataset size and model complexity.

## Case study

As our results in d) will demonstrate, the DNN-SSF embedding architecture is able to infer inter-individual variability in nonlinear responses, and this variability can be visualized in embedding space, where individuals that are closer to each other in the embedding space are more similar in their movement predictors. To show the practical use of this, we applied the method to real animal trajectories to explore the inter-individual variability in the movement of eight wild boars (*Sus scrofa*), with individuals inhabiting the urban area of Berlin, Germany's capital, and rural surroundings of the federal state of Brandenburg. In the original study (Stillfried et al. 2017), wild boars were tracked at 30 min intervals and grouped into urban and rural individuals based on their vicinity to the urban matrix to study resource selection. Importantly, the grouping was *a priori* and not based on their movement behavior. This allows us to test if clustering individuals by their movement behavior would match the home range-based grouping.

The data were subsampled to a time resolution of 720 minutes, following Signer et al. (2025). For each observed step, we randomly drew 20 steps from the empirical gamma distribution of step lengths (sl_) and an empirical distribution of the turning angles (ta_), which was estimated using the R package 'PDFEstimator' (v4.5, Farmer and Jacobs 2018), and compared them to their observed step. We used the distance to water (dtw), refuge habitat (hab; percent tree cover per raster cell) and impervious surface (imp) of the urban matrix as covariates (for further information on environmental covariate preparation see Signer et al. 2025).

We used permutation importance to infer the importance of the selection and movement variables, as well as the pair-wise importance of their interactions. Effects were visualized using accumulated local effect plots.

## Results

## For linear effects DNN-SSFs show low bias and well-calibrated p-values and confidence intervals

Our first scenario was designed to ensure that DNN-SSFs can learn simple linear effects of the environment on the movement, and that p-values and confidence intervals calculated from the DNN using our xAI methods are well calibrated. Our results show our DNN-SSF accurately recovers the linear effect of the environment across a sensible range of effect sizes (Figure 2). Simulated effect sizes ranged from -2 to 2 for a standardized predictor, which amounts to strong effects considering the logit link, showing that the DNN delivers near-unbiased effects across the entire possible range of effect sizes. The Type I error rates for both DNN-SSFs and GLM-SSFs were close to the nominal 0.05 level, as expected from a perfectly calibrated model. The coverage, i.e. the proportion of true slopes lying within the 95% confidence interval of the estimates, was also close to the expected 0.95 level. The only performance issue we found in comparison to the GLM-SSF is that effect estimates of the DNN-SSFs showed a slight, albeit non-significant, bias (difference between the true and estimated slope) towards zero for strong effects (see Figure 2b). This means that we see (non-significant) hints that the DNN shows a slight regularization bias that pushes effect estimates towards smaller values in the absence of overwhelming data, which would be in line with existing research (e.g. Pichler & Hartig, 2023a).

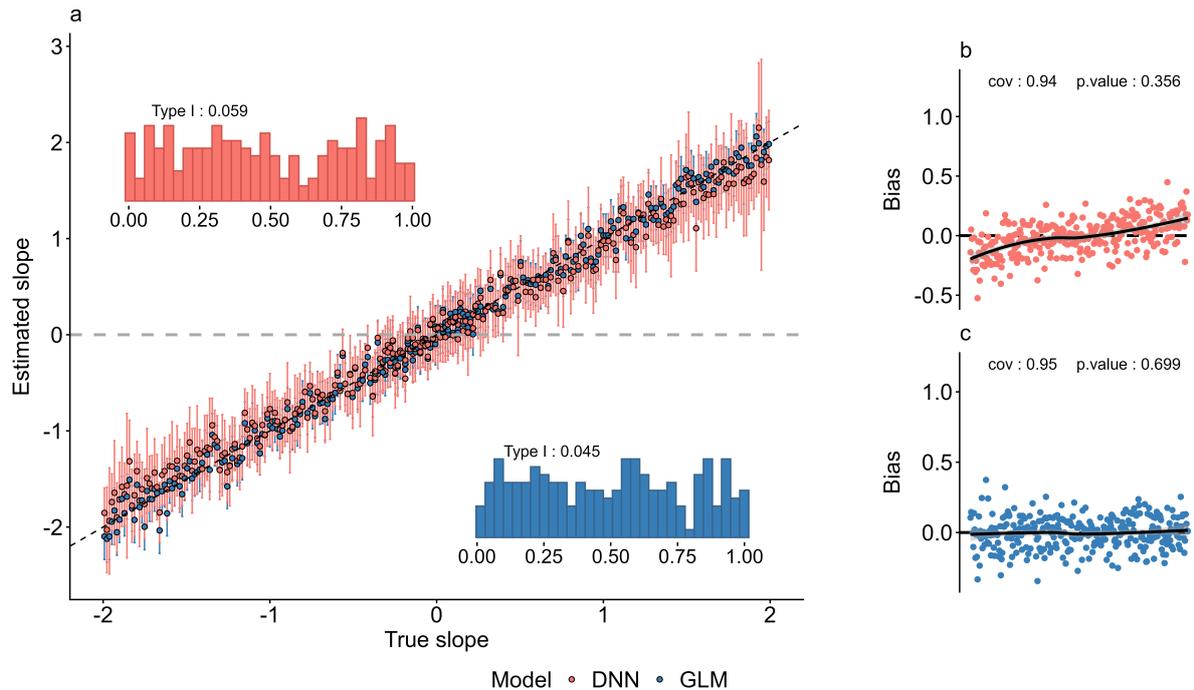

**Figure 2: Validation of DNN effect estimates for a linear response to habitat:** we compare slope estimates between DNN-SSF (red) and GLM-SSF (blue) models. Data simulated with a linear response to one predictor reflecting a range of possible strengths of the habitat preference from -2 to 2. a) Each point represents the mean estimated slope across simulations for a given true slope (ranges from -2 to 2), with vertical error bars indicating 95% confidence intervals. The diagonal black dashed line represents the true slope, which serves as a reference baseline. Histograms of p-values under the null hypothesis (true slope = 0) are shown above and below the main panel for the DNN and GLM models, respectively. b and c) Bias (difference between the true and estimated slope), coverage (true slope within the 95% confidence interval), abbreviated as 'cov', and the corresponding p-value for DNN-SSF (b) and GLM-SSF (c). X-axis of b & c is the same as the one on a. We fitted a GAM to check whether the bias is significantly different from zero. Although there is a slight pattern in the GAM fit for the DNN (b), the p-values for both bias checks were not significant.

## DNN-SSFs are as good as GAM-SSFs at recovering nonlinear effects

Our second scenario aimed to check whether DNNs can correctly infer the shape of nonlinear environmental effects on the movement. We simulated two scenarios, one with a relatively simple hump-shaped response that could arise from a niche-type environmental preference (Figure 3a), and one scenario with a more complicated wiggly response (Figure 3b). Both the DNN-SSFs and GAM-SSFs were able to accurately approximate these functional responses (Figure 3). Although subtle in the visual representation, the mean squared error (MSE) was slightly lower for the DNN-SSF compared to the GAM-SSF, which may be connected to the fact that the DNN-SSF was better able to recover the functional complexity of the true underlying curve, whereas the estimated effect of the GAM seems a bit too "wiggly".

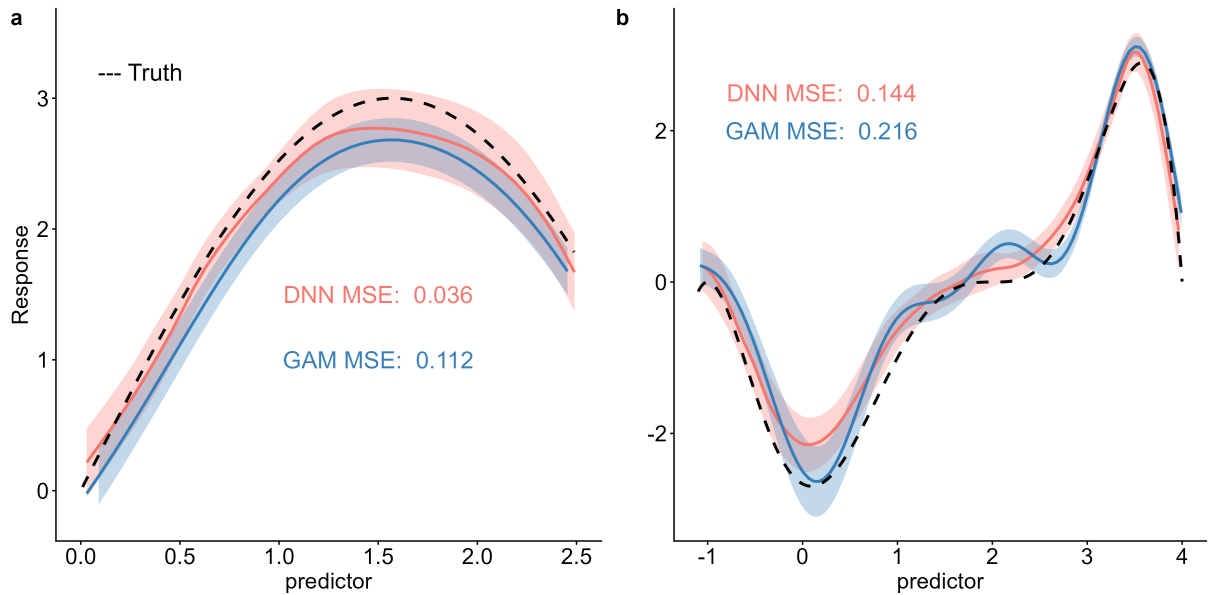

**Figure 3: Inferring nonlinear effect (smoothing):** A comparison of a DNN-SSF (red) and a GAM-SSF (blue; with a smoothing term on the predictor) when recovering two nonlinear functional relationships between a single predictor and the movement from simulated data. The true shape of the functional response is represented by the black dashed line, while the colored lines correspond to the mean estimates of the two models. The shades represent the 95% confidence intervals of the mean estimates, and the mean squared errors (MSE) are indicated in red and blue.

## DNN-SSFs are good at recovering statistical interactions between many predictors

Our third scenario was designed to compare DNN-SSFs and GLM-SSFs in their ability to infer linear main effects and linear interactions from a larger number of predictors (9 predictors, 9*8/2 = 36 possible interactions). The challenge in this setting is not so much the complexity of the response, as in the previous scenario, but the comparatively large number of predictors and their interactions.

Our results show that both models were able to accurately identify the true main effects, as indicated by a high alignment of the circle sizes on the diagonal with the respective colored point sizes and by the low MSE values (Figure 4). We do not find pronounced differences between both approaches.

For the interactions, both models identified the three true interactions as important predictors. The DNN-SSF tended to slightly underestimate the effect of the true interactions and thus exhibited larger MSE and bias for these values (Figure 4); whereas the GLM-SSF tended to show more variance (Figure S1) and larger MSE (Figure 4) in effect estimates for interactions that were zero. This pattern is in line with the expectation that a DNN creates a shrinkage bias (Pichler and Hartig 2023a), which means that the model tends to push all interaction estimates towards smaller values and thus trades off bias for reduced variance.

Averaged across all effects, the MSE was 0.08 for the DNN-SSF and 0.14 for the GLM-SSF, showing that the DNN-SSF outperforms the GLM-SSF regarding the average estimation error. This observation is compatible with the general insight in machine learning that in situations with many predictors and few data, a shrinkage bias as identified above often leads to more stable parameter estimates (Hastie et al. 2009).

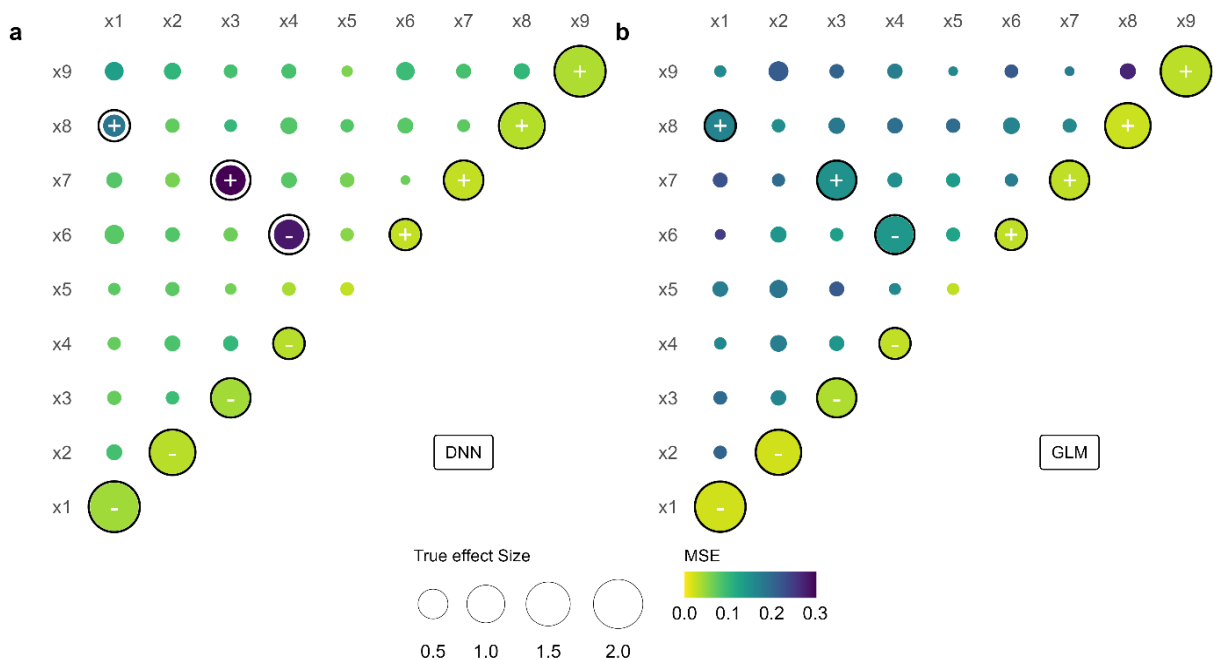

**Figure 4: Inferring effects of nine predictors with multiple interactions:** The figure visualizes the estimated main effects (diagonal) and two-way interactions of nine predictors (x1 to x9), comparing performance between a DNN-SSF (a, left panel) and a GLM-SSF (b, right panel). The data were simulated with main effects increasing in both directions from x5, which itself had no effect, and interactions between x1-x8, x3-x7 and x4-x6, indicated by black circles at the respective intersections. The size of each black circle reflects the true effect magnitude, while the size of the overlaid colored disk represents the estimated effect.

Complete overlap between the black circle and the colored disk therefore indicates unbiased estimation. Color represents the mean squared error (MSE) of the estimated effect. Averaged across all effects, the MSE was 0.08 for the DNN-SSF and 0.14 for the GLM-SSF. For the DNN-SSF the average MSE was 0.27 for the true interactions, 0.08 for null interactions, and 0.03 for the main effects. Corresponding values for the GLM-SSF were 0.15, 0.18, and 0.02. The sign information ("+" or "−") inside circles show the direction of the true effect. The DNN was trained with a learning rate of 0.01 for 150 epochs.

## DNN embeddings can detect inter-individual differences in environmental preferences

In our fourth scenario, we simulated data for 20 individuals that were divided into four groups, with each group responding differently to the environmental predictors, akin to four different animal "personalities".

Our results show that a DNN-SSF with an embedding layer correctly retrieves these four groups, which manifests by individuals from the same group clustering in the embedding space (Figure 5). Furthermore, through back-projecting the influence of the embedding position on the covariate effects, we find patterns that correspond with our simulated processes: Covariates with no effect on the movement ($\beta = 0$) are not correlated with the embedding position of the individuals (Figure 5, black arrows). For predictors without collinearity, the embedding positions have varying but uncorrelated effects on the habitat selection (Figure 5, green arrows). For the group of collinear predictors, the embedding position has similar effects on the habitat selection, as one would expect (Figure 5, violet arrows) because these predictors were generated as variations of the same underlying environmental gradient.

Overall, our results show that the embedding structure was successful in identifying individuals with similar behavior or personality, and back-calculating the effect of the embedding on the movement behavior allows interpreting the embedding position of individuals in terms of their behavioral differences in a meaningful way.

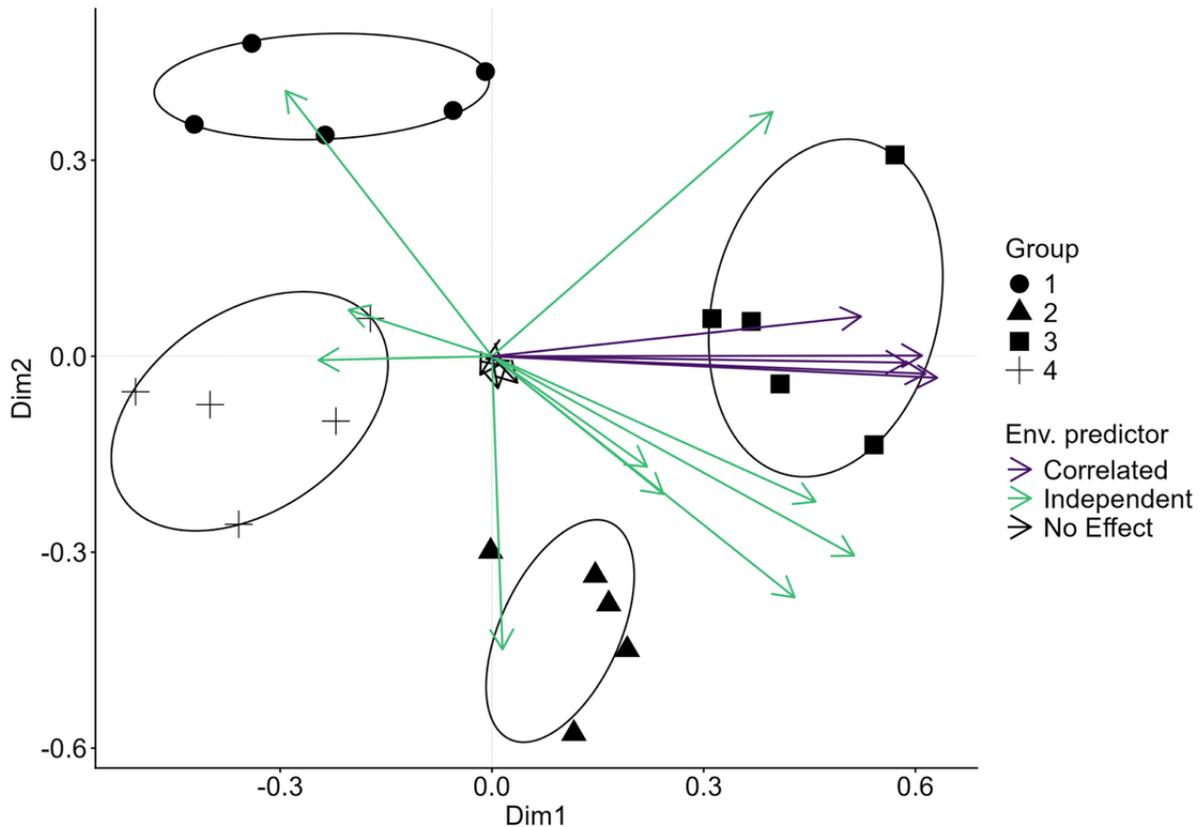

**Figure 5: Individual responses of the focal individuals to environmental predictors**. The embedding space depicts the projection of the response of 20 individuals to 20 environmental predictors. The shape of the points indicates the four different groups of individuals. The ellipses indicate group-level clustering within the embedding space, capturing the within-group similarity in selection patterns. Ellipses are added for visualization and do not represent the results of a formal clustering analysis. Arrows represent the effect of the embedding position on the selection coefficients of the individuals. They are color-coded by type: independent of each other (green), correlated (violet), and those with no effect (black). The length of an arrow reflects the magnitude of the predictors effect, with longer arrows indicating a stronger effect.

## DNN-SSF embeddings can detect different classes of opponents

In our fifth scenario, we aimed to evaluate whether DNN embeddings can detect differences in how opponents affect movement of focal individuals. We therefore applied the embedding layer to the opponents rather than the focal individuals and simulated a scenario with 3 types of animals that act 1) neutral, 2) attractive, or 3) repellent on all other individuals. The results reveal again a clear grouping pattern within the embedding space (Figure 6), indicating that the model effectively captures variations in the opponents' effect on focal individuals' movement. Moreover, the positioning of the groups relative to the effect of the embedding position (arrow in Figure 6) is easily interpretable and in line with our simulations: the neutral group is placed in the center, and the repelling group selects more positively and the attracting group more negatively for distance to the opponent, as indicated by the arrow.

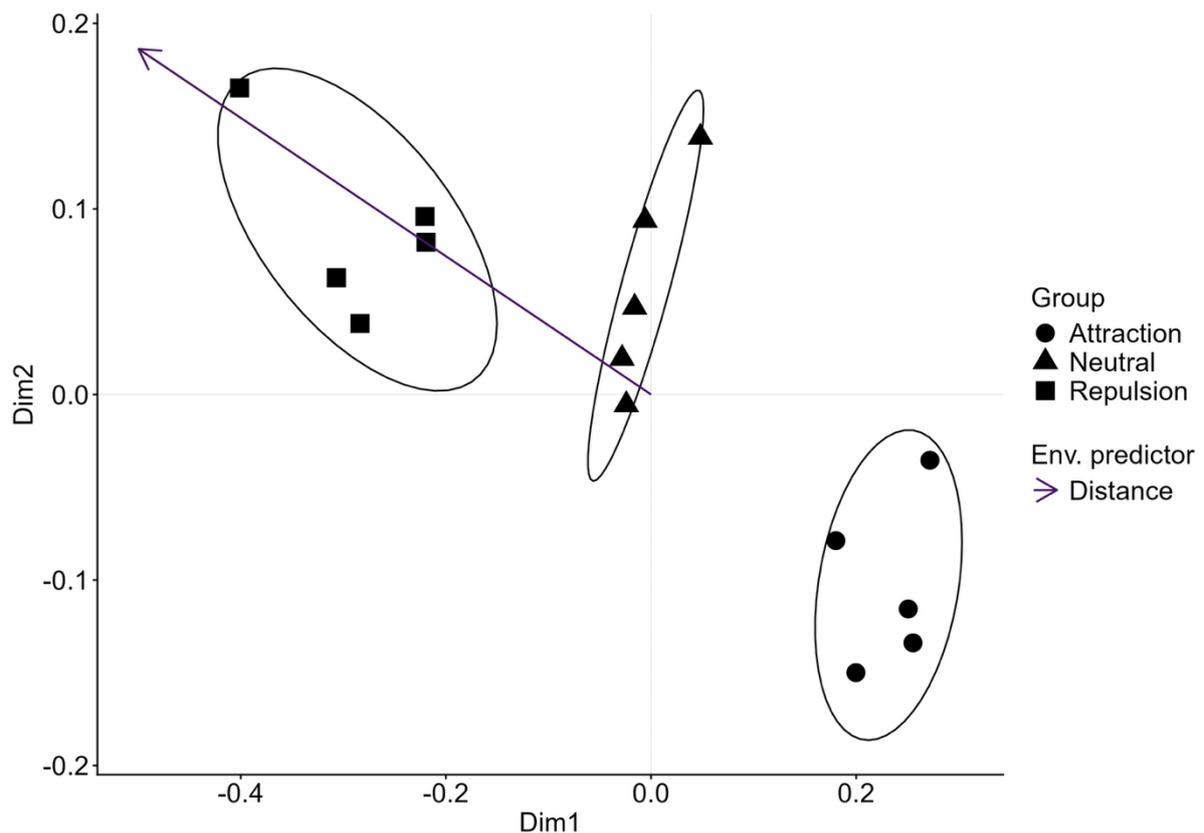

**Figure 6: Effects of individual opponents on the movement of the focal individual.** The embedding space depicts the projection of the effects of 15 opponent individuals (nearest individual to the focal individual) on the focal individual. The shape of the points indicates the three different groups of opponents. The effect of the embedding position on the selection coefficient is depicted by an arrow. In this example, the only predictor of the movement is the distance to the opponent, so there is only one arrow. Note that the positioning of the groups relative to the distance arrow is in line with our simulations: the neutral group in the center, the repelling group (e.g. prey fleeing a predator) selects positively for distance, and the attracting group (e.g. social group movement) selects negatively to distance.

## Case study

As a final case study, we applied a DNN-SSF with embedding architecture to data of eight wild boar (Stillfried et al. 2017) at a resolution of 720 min. Using permutation importance, we quantified the influence of the five predictors on movement decisions of wild boars. Distance to water (dtw) and imperviousness showed the highest overall importance (Figure 7a). Turning angle (ta_) appeared as the third-most important predictor. This is interesting because, having sampled the random steps from a non-parametric kernel estimator, we would not necessarily have expected large turning angle effects (which essentially correct the initial estimate). An analysis of pair-wise importance (Figure 7b) shows that these values mostly originate from interactions of the turning angle with distance to water and imperviousness, suggesting that the directionality of the movement changes with habitat context. This finding underlines a distinct benefit of the DNN-SSF, as such movement-habitat interactions are difficult to implement in conventional SSF models.

For distance to water, the most important variable, we found nonlinear negative effects that differ only slightly among individuals (Figure 7c). Examining differences in the embedding position of the individuals, we found that rural wild boars form a distinct cluster, suggesting that they show similar habitat selection. The cluster position suggests that rural individuals showed a stronger selection for habitats farther from water compared to urban individuals. (Figure 7d). Urban wild boars differ more strongly in their movement behavior. Behavioral differences are mostly explained by turning angle and imperviousness (Figure 7d). These results overall support the hypothesis that urban wild boars have adjusted their behavior to their urban habitat, but there seems to be a stronger variation of behavior between urban individuals, possibly related to stronger landscape heterogeneity and disturbances in urban habitats.

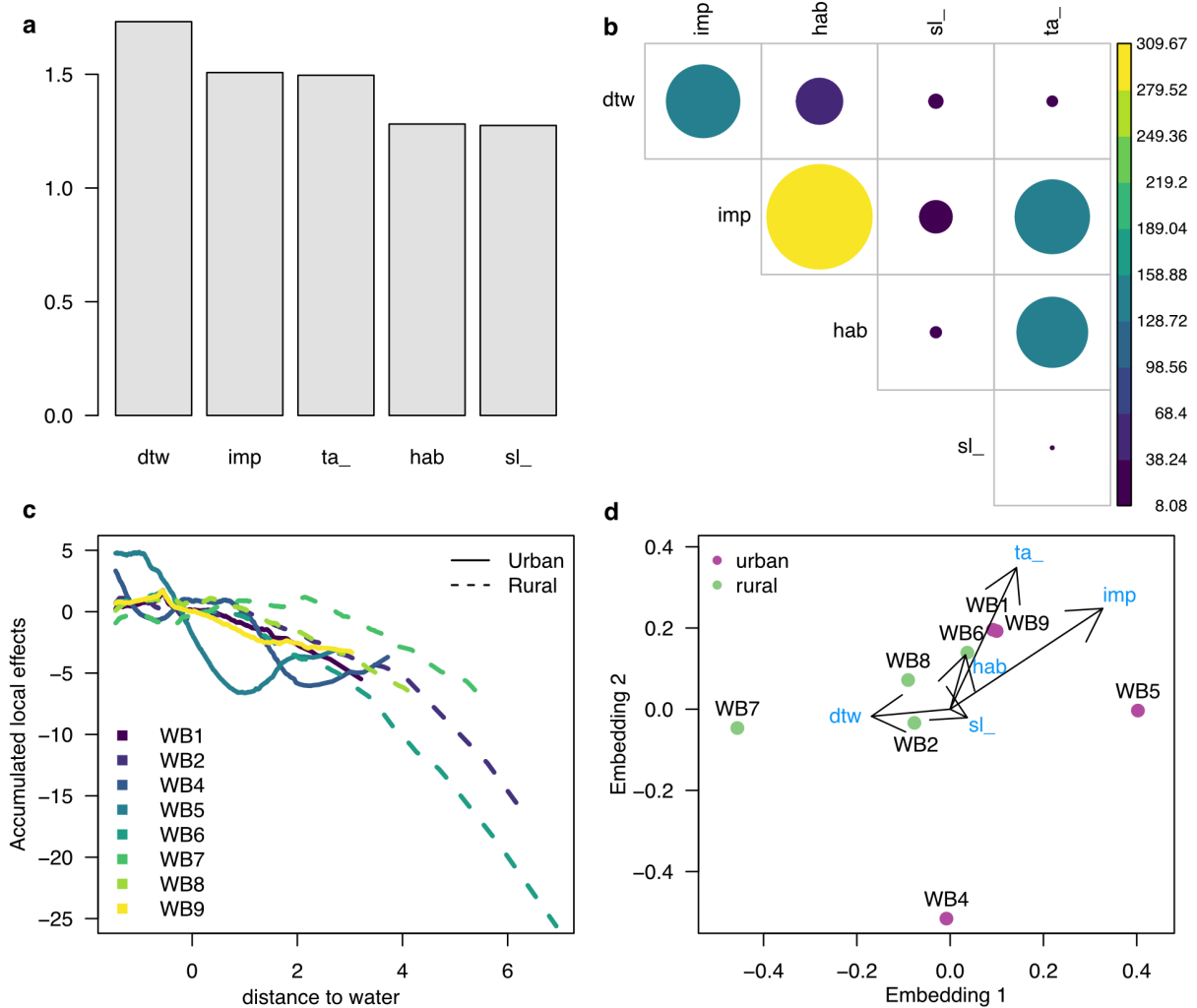

**Figure 7: Analysis of wild boar data (Stillfried et al. 2017) with a DNN-SSF**. Eight wild boars were tracked in Berlin (urban) and in its surrounding rural area. We estimated a DNN-SSF simultaneously for all eight individuals, that were subsampled to a time resolution of 720 minutes (following Signer et al. 2025) with turning angle ("ta_"), imperviousness ("imp"), distance to water ("dtw"), step length ("sl_"), and % tree cover as refuge habitat ("hab") as predictors. The selection effects of these predictors are modified by an embedding layer in the DNN. Looking at the permutation importance of the predictors (a), distance to water and imperviousness had the strongest effects. The interaction importance showed strong interactions between the selection variables (imperviousness and habitat), as well as between the selection variables and the movement, particularly the turning angle (b). Accumulated local effect plots of the distance to water (c) show a nonlinear decaying effect of the predictor on the selection choice. Behavioral differences between rural individuals were smaller than between urban individuals, with distance to water explaining the main difference between urban and rural individuals and turning angle and imperviousness explaining variation between urban individuals (d).

## Discussion

The goal of our study was to evaluate the performance of fully connected deep neural networks as a flexible curve fitting structure within the SSF framework. Our results show that DNN-SSFs can be a powerful extension of the step selection framework. They compare well against alternative models such as GLM-SSF or GAM-SSF, especially for capturing complex nonlinear selection patterns with predictor interactions and inter-individual variability. In particular, we showed that DNN-SSFs can correctly retrieve linear effects, including confidence intervals and p-values (Figure 2) as well as nonlinear effects of single predictors similar to a GAM (Figure 3); perform well at inferring sparse interaction structures between multiple predictor variables (Figure 4) and allow model differences between data groups (e.g. individuals) in an embedding structure that acts similar to a nonlinear random effect on all predictors, but offers an easy interpretation akin to an ordination biplot that shows the similarity effects between individuals or groups (Figures 5, 6). Moreover, the DNN-SSF naturally considers interactions between habitat and the movement kernel (Figure 7b), which seems an important advantage for the analysis of empirical movement data.

Our results complement existing pioneering studies on the use of deep learning (DL) models in movement ecology that concentrated more on the benefits of DL for using complex predictors (Cífka et al. 2023; Forrest et al. 2025b) and less on their inference capabilities. Our results suggest that the DNN can retrieve nearly unbiased effect estimates that are in line with Pichler and Hartig (2023a). We also find that effect estimates from DNN can achieve acceptable Type I error rates and coverage of confidence intervals. This shows that at least simple DNNs are not a completely different model class but seamlessly extend previous GLM- or GAM-based step-selection models for inferring habitat selection.

The benefit of DNN-SSF over existing approaches is arguably most relevant when we are faced with complex data and responses. Our results show that in simulated data, the DNN-SSF was able to retrieve linear two-way interactions as well as a GLM-SSFs. The advantage of DNN-SSFs, however, is that these interactions did not have to be defined *a priori*, nor would they have to be linear. This becomes particularly evident when habitat selection variables interact with movement variables. These interactions are usually not assumed, as it is likely just not feasible given the combinatorially large number of interactions; however, in our case study, we found large interaction importance (Figure 7b) between turning angle and habitat variables, highlighting that these interactions should be given more focus in the future. We also showed that the DNN-SSF was able to recover simple nonlinear functional forms as well as or better than a GAM-SSF. These results suggest that DNN-SSFs, due to their flexibility and ability to approximate complex functions, may aid ecologists in automatically inferring the complex functional relationships from the data. For instance, daily movement patterns of animals often exhibit complex, nonlinear patterns that can be influenced by a variety of factors such as temperature (Thaker et al. 2019), time of day (Klappstein et al. 2024), presence of predators (Fischhoff et al. 2007), or terrain (Avgar et al. 2013). Although we only tested nonlinear relationships with one predictor, we anticipate that the DNNs excel particularly for a large number of nonlinear interactions. So far, those could only be specified using tensor splines in a GAM-SSF, but these structures are very costly and arguably less robust.

A downside of the flexibility of DNN-SSFs is a potential loss of power. Model flexibility has therefore been balanced against the available data, based on the bias–variance trade-off (James et al. 2013). As model complexity increases, effective regularization becomes essential to control variance, but such regularization and optimization choices introduce small amounts of bias into the estimates (Zou and Hastie 2005). Although our study does not focus on these aspects of the modelling process, the 'citoMove' R package inherits many tunable regularization strategies (e.g., dropout, L2, and L1) from the 'cito' R package that help researchers control the bias–variance trade-off (Amesoeder et al. 2024b).

An additional advantage, which is to our knowledge, highlighted for the first time in this study, is that DNNs provide interesting options to specify nonlinear random effects through an embedding layer linked to the group identity. These structures can be seen as generalizations of random-slope models (Dingemanse and Dochtermann 2013; Hertel et al. 2020; Chatterjee et al. 2024; Klappstein et al. 2024) or hierarchical GAMs (Pedersen et al. 2019) that were previously used to analyze variation in movement behavior among individuals. In the embedding approach, each individual or group is placed in an n-dimensional embedding space during model training, and the location in this space then codes the type of environmental response, with individuals that are placed close to each other having similar (but potentially complex) environmental responses. The embedding position thus simultaneously encodes linear effects, nonlinearities, and interactions in one vector, and the similarity of groups can then be visualized similar to an ordination biplot.

Using this approach, we showed that DNN-SSFs can effectively capture inter-individual variability in habitat selection effects. It is assumed that such among-individual variation in habitat-selection strategies is not only driven by external factors (such as habitat quality, predation risk, seasonal condition, etc.) but also arises due to differences in personality or behavioral state, experiences in the past, or physiology (Hertel et al. 2020). Accounting for such variation has long been recognized as essential for understanding the movement of animals (Shaw 2020), as it shapes predator-prey interactions (McGhee et al. 2013), dispersal (Clobert et al. 2009; Cote and Clobert 2010), fitness (Smith and Blumstein 2008) and thus population dynamics (Del Mar Delgado et al. 2018; Shaw 2020). Our results show that DNN-SSFs offer a very appealing analytical pipeline to detect such differences.

Another interesting application of the embedding architecture is to reveal similarities in the effect an individual has on other individuals, which could help, for example, to study social structures (Langrock et al. 2014), responses in predator-prey systems or among competitors (Vanak et al. 2013) and group dynamics (leadership or following dynamics - Strandburg-Peshkin et al. 2018). In a simulated case study, we showed that the DNN-SSF can separate groups of individuals that act attracting, repellent, and neutral on other individuals. Existing methods, such as the framework by Schlägel et al. (2019), also allow to estimate the effect of an individual on the movement behavior of other individuals. However, this approach relies on fitting an SSF for each individual independently, whereas our DNN-SSF embedding architecture allows to consider the entire group at once. We believe that the approach presented here may help to uncover important mechanisms contributing to migration/dispersal, home-range analysis, group dynamics, or foraging strategies (Worton 1989; Jeltsch et al. 2013). Although not tested here, this framework offers a novel opportunity to incorporate statistical interactions between the group of the opponent and environmental predictors. This enables analysis of how the habitat selection of a focal individual varies depending on the group of nearby individuals.

A purposeful limitation of our study was that we concentrated on fully connected neural networks, also known as multi-layer perceptrons (MLP), which is the simplest neural network architecture. An obvious extension of our work, which was already considered in other papers, would be to extend the neural networks to more advanced deep learning architectures. For example, convolutional layers could enable the direct use of raw spatial imagery (e.g., LiDAR or satellite image) as movement predictors (Forrest et al. 2025b), rather than their summary statistics as predictors of the movement. Another option would be the use of recurrent neural networks or attention-based architectures that include the past movement as context, thus in some sense creating an internal state of the animal that affects the movement decision (Cífka et al. 2023). These extensions represent highly promising avenues as DL continues to evolve, offering new opportunities for combining ecological inference with the flexibility of modern ML. However, in this study, we concentrated on the simpler MLPs so that the same predictors could be used and thus compare model performance to traditional SSF models.

## Conclusion

Deep Learning approaches offer a promising alternative to existing step-selection modeling frameworks for understanding drivers behind animal movements. In this study, we demonstrated that by inserting simple fully-connected DNNs within the familiar step-selection framework, we obtain a model that has similar properties to a GAM-SSF, in that it can infer nonlinear relationships while retaining interpretability, p-values and confidence intervals. However, this new model offers greater flexibility than GAM-SSF regarding the specification of (nonlinear) interactions and provides exciting opportunities to analyze inter-individual differences in selection effects (e.g. animal personality) and in the effect of individuals on other individuals. Moreover, once we are already in a deep learning framework, it is relatively straightforward to extend the modelling approach to more complicated network architecture such as CNNs or transformers. To facilitate the uptake of the methods described in this paper, we provide them in the R package 'citoMove'.

## Software

All analyses and simulations were performed in R 4.5.2 (R Core Team 2025) with the packages 'citoMove', 'survival' (Therneau 2024), 'amt' (Signer et al. 2019), and 'mgcv' (Wood 2025).

## Acknowledgements

This work was supported by the German Research Foundation (DFG) Research Training Group "BioMove" (DFG-GRK 2118/2). We thank Milena Stillfried for collecting the wild boar field data. We thank Aastha Tapaliya for the support with the preliminary exploration of how machine learning can be used to infer movement interactions.

## Data Availability

Upon acceptance of the manuscript, we will make the code to reproduce the analysis presented in this paper available as a GitHub repository and produce a persistent snapshot of this repository via zenodo. The citoMove R package will be made freely available at latest upon acceptance of the manuscript, ideally via CRAN.

## Conflict of Interest statement

The authors have no conflicts of interest to declare.

## Author Contributions

Thibault Fronville, Maximilian Pichler, Viktoriia Radchuk, and Florian Hartig conceived the ideas and designed methodology; Thibault Fronville simulated the data; Thibault Fronville and Maximilian Pichler analyzed the simulated data. Marius Grabow and Stephanie Kramer-Schadt prepared the spatial data for the case study; Maximilian Pichler analyzed the field data; Thibault Fronville, Maximilian Pichler, Florian Hartig, Viktoriia Radchuk and Johannes Signer contributed to the development of the citoMove R package. Thibault Fronville, Maximilian

Pichler, Viktoriia Radchuk, and Florian Hartig led the writing of the manuscript. All authors contributed critically to the drafts and gave final approval for publication.

## Supplementary Materials

Analyzing animal movement using deep learning

## S1. Simulation settings

In our study, we simulated movement using a step-selection framework in which at each time step, an individual selects its next location from a set of randomly generated available steps, with selection driven solely by environmental covariates.

At each step, we generated a set of covariates for all the available steps. For each of the environmental covariates we sampled random values from a uniform distribution:

$$x \sim U(0,1) \quad \text{equation S1}$$

To enhance interpretability and reduce collinearity in the presence of interaction terms, we centered the covariates prior to their usage in the selection model.

To get the selection probability $p(i)$ we first need to calculate the selection score for each available step:

$$w_i = \exp\left(\sum_j \beta_j x_{ij} + \sum_k \gamma_k x_{i,pk} x_{i,qk}\right) \quad \text{equation S2}$$

where $x_{ij}$ is the j-th covariate value for the i-th available step and $\beta_j$ is their corresponding selection coefficient. $\gamma_k$ is the coefficient for the $k$-th interaction between covariates $x_{pk}$ and $x_{qk}$. $w_i$ corresponds to the relative habitat selection strength (also called unnormalized weights) of the available step $i$. Nonlinear relationships can be incorporated by including transformed covariates among the predictors $x_{ij}$.

The selection probability for step $i$ is then obtained by normalizing the weights for each step $i$ over all the available weights:

$$p(i) = \frac{w_i}{\sum_{i=1}^n w_i} \quad \text{equation S3}$$

Finally, the true next step was selected by drawing from a multinomial distribution with probabilities $p(i)$. This ensures that the steps with higher selection strengths are more likely to be chosen, while allowing stochastic variation in the selection process.

## S2. DNN-SSFs are good at recovering statistical interactions between many predictors

Our third scenario was designed to compare DNN-SSFs and GLM-SSFs in their ability to infer linear main effects and linear interactions from a larger number of predictors (9 predictors, 9*8/2 = 36 possible interactions).

Our results show that both models were able to accurately identify the main effects, as indicated by a high alignment of the circle sizes on the diagonal with the respective colored point sizes and by the low Variance values (Figure S1). We do not find pronounced differences between the two modeling options.

For the interactions, both models identified the three true interactions as important predictors. The DNN-SSF tended to slightly underestimate the effect of the true interactions (circles are not filled), while the GLM-SSF tended to show more variance (Figure S1) in effect estimates for interactions. This pattern is in line with the expectation that a DNN creates a shrinkage bias, which means that the model tends to push all interaction estimates towards smaller values and thus trades off bias for reduced variance.

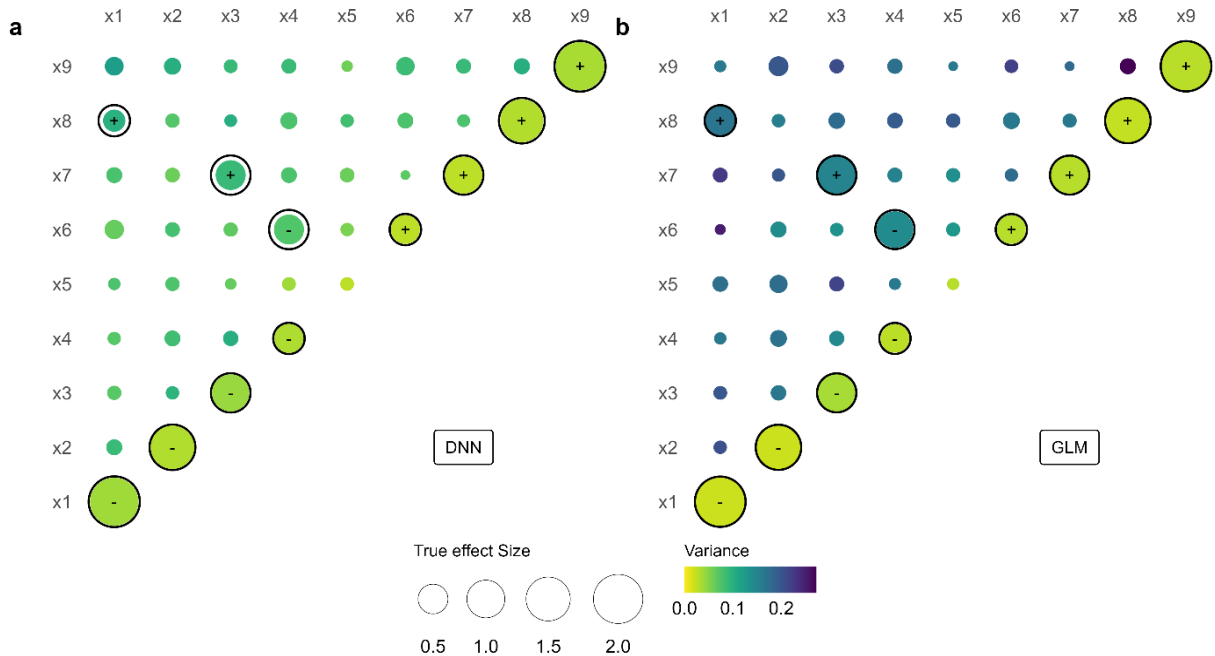

**Figure S1**: A matrix-based visualization of pairwise interaction effects among predictors (x1 to x9), comparing performance between a DNN-SSF (left panel) and a GLM-SSF (right panel). The diagonal represents the main effect of the 9 predictors. The other cells represent a specific interaction term between two predictors. Interactions were present between x1-x8, x3-x7 and x4-x6. The true effect size is indicated by the size of the black circle while the estimated size is indicated by the size of the colored points. The variance of the estimated effect is indicated by color. Sign information ("+" or "−") inside circles show the direction of the true effect.